\documentclass[superscriptaddress,
 reprint,
 amsmath, amssymb,
 aps, preprintnumbers, nofootinbib, floatfix, longbibliography
]{revtex4-2}

\usepackage{graphicx}
\usepackage{dcolumn}
\usepackage{bm}
\usepackage{hyperref}
\usepackage[margin=1in]{geometry}
\usepackage{amsfonts}
\usepackage{amsmath}
\usepackage{amssymb}
\usepackage{xcolor}
\usepackage{bbold}
\usepackage{tikz}
\usetikzlibrary{quantikz2}

\begin{document}

\title{Three-flavor supernova neutrino simulation using a hybrid quantum-classical algorithm with qutrits}

\author{Daniel J. Heimsoth}
\email{dheimsoth@wisc.edu}
\affiliation{Department of Physics, University of Wisconsin--Madison, Madison, WI, USA 53706}

\author{Pooja Siwach}    
\affiliation{Department of Physics, University of Arizona, Tucson, AZ, USA 85721}
\affiliation{Facility for Rare Isotope Beams, Michigan State University, East Lansing, Michigan 48824, USA}

\author{A. Baha Balantekin}
\email{baha@physics.wisc.edu}
\affiliation{Department of Physics, University of Wisconsin--Madison, Madison, WI, USA 53706}

\date{\today}

\begin{abstract}
We simulate a self-interacting three-flavor neutrino system within a core-collapse supernova using a hybrid classical-quantum algorithm on a qutrit computer. Based on the Dirac-Frenkel evolution equations, we employ a variation of the quantum-assisted simulator to calculate the system's time evolution operator by performing qutrit Hadamard tests to find expectation values of unitary operators in the Hamiltonian. The time evolution simulation is then done classically. We find that the hybrid algorithm produces results comparable to an exact numerical integration out to times of $t \approx 30 \,\omega_0^{-1}$ with time step $\delta t = 0.005 \,\omega_0^{-1}$, where $\omega_0$ is the energy scale of the single neutrino vacuum oscillations. We discuss the lessons learned in simulating neutrino systems using this hybrid quantum-classical algorithm, along with the advantages it offers over quantum Trotterization. 
\end{abstract}

\maketitle

\section{Introduction}

The realization of low-noise, scalable quantum computers promises to allow the computation of complex problems that classical methods struggle with, either due to processing, memory, or time limitations. These quantum machines are still on the mid- to far-term horizon, however, leaving current research efforts with noisy, intermediate-scale quantum (NISQ) devices. These quantum computers, while useful in solving various types of problems of interest to scientists across fields, struggle with relatively low fidelities on entangling gates, the very operations that give quantum computers their true power over classical methods. Thus, it is advantageous to limit the depth of quantum circuits to minimize the number of entangling gates applied in succession.

One method to achieve this is to actually \textit{not} use the quantum computer, at least in portions of the calculation that a classical computer can solve efficiently. The quantum computer is used only where it provides some scaling or processing advantage over classical methods, in particular for calculating matrix elements of unitary operators. These hybrid classical-quantum algorithms have been relatively popular in the era of NISQ devices to circumvent their limitations due to poor entangling gate fidelities and issues scaling up the number of qubits~\cite{Cerezo:npj2022,Yuan2019}.

Time evolution of physical systems that easily map onto spin states is a prime candidate for such a hybrid calculation, since classical computers are very good at calculating time-evolved states (through matrix multiplication) and quantum computers are capable of calculating expectation values of spin operators at fairly low circuit depth. {\color{black} Collective neutrino flavor oscillations are such a system. (For reviews see ~\cite{Duan:2010bg,Chakraborty:2016yeg,Balantekin:2018mpq,Tamborra:2020cul,Volpe:2023met,Balantekin:2023qvm,Johns:2025mlm}). In collective neutrino oscillations a single neutrino in the two-flavor approximation is a realization of SU(2) and thus maps directly onto a qubit state. Therefore, the collective many-body neutrino oscillation phenomenon beyond mean-field limit has been extensively studied with both classical~\cite{cervia:2019, Cervia:2022pro, Lacroix:2022krq,Patwardhan:2022mxg,Illa:2022zgu,Martin:2023gbo,Martin:2021bri,Martin:2023ljq,Lacroix:2024pbb,Carlson:2026mir,Neill:2024klc,Chernyshev:2024pqy,Bhaskar:2024myw,Kiss:2025jgt,Laraib:2025uza,Laraib:2025ziz} and quantum simulation methods~\cite{Yeter-Aydeniz:2021olz, Amitrano:2022yyn, Hall:2021rbv,Siwach:2023wzy,Balantekin:2023qvm,Turro:2024shh} but mostly under the two-flavor approximation}. From experiment, however, we know that there are (at least) three active flavors of neutrinos, and the oscillations among them live in SU(3). Pauli operators are substituted with Gell-Mann matrices, two energy levels become three, and {\color{black} the algebraic structure of the system becomes more complicated}~\cite{siwach:2023prd}. Thus, the natural quantum device for studying three-flavor neutrino states is instead the three-level qutrit~\cite{Balantekin:2024pwc}, which brings complications to the calculation that do not show up in the qubit case~\cite{Mangin-Brinet:2026prd,Spagnoli:2025prd,Chernyshev:prr2025}. Our goal in this paper is to identify and overcome these obstacles while simulating the time evolution of interacting three-flavor neutrinos in a core-collapse supernova (CCSN) environment.

In the following section, we introduce the mathematical framework that we will use to study supernova neutrino interactions. Then, in Sec.~\ref{sec:qutrit_algo}, we translate our Hamiltonian onto a collection of qutrits, three-level quantum states that match the behavior of three-flavor neutrino oscillations, and describe the process of implementing the Dirac-Frenkel quantum-assisted simulator. We present our results from a Dirac-Frenkel hybrid calculation in Sec.~\ref{sec:results}, along with a quantification of errors from this approach. We follow with Sec.~\ref{sec:discussion}, a discussion of the benefits, drawbacks, and future developments of this hybrid algorithm in the context of the current and mid-term quantum computing landscape.

\section{Supernova Neutrinos} \label{sec:supernova}

Neutrino oscillations have a very important role in determining the dynamics and observables of core-collapse supernovae, as neutrinos are the main source of energy transport both within and out of the collapsing star~\cite{Martinez-Pinedo:2017ksl}. They are predicted to dictate nucleosynthesis rates, electron fraction, and protoneutron star mass, among other quantities~\cite{Balantekin:2023ayx}. Because of their relatively weak interaction with the surrounding matter, neutrinos are the only direct probe of dynamics in the core of the collapsing star. The nature of their flavor oscillations has a {\color{black} nontrivial} effect on supernova processes, as only electron neutrinos (and antineutrinos) have charged current interactions with the large mass of electrons in the core and surrounding matter. {\color{black} Further, these flavor oscillations affect the detectability of supernova neutrinos in Earth-based experiments since different detection channels are sensitive to different neutrino flavors~\cite{choi:2025, Mirizzi:2015eza, Pitik:2022}.}

Modeling neutrino interactions within a CCSN is a nontrivial task, as on average a CCSN is predicted to produce around $10^{53}$~erg in neutrinos which interact both with themselves and with the surrounding matter~\cite{Fuller:2022nbn}. Further, the strength of the neutrino-neutrino interaction depends on the angle of the incoming particles, neutrino energy, and the neutrino density, all of which may be different for every interaction. In order to make this problem more tractable, some averaging is required; commonly, the single-angle approximation and neutrino bulb model are used~\cite{Duan:2006an,Qian:1994wh, Bell:2003mg, Friedland:2006ke}. The single-angle approximation assumes that the flavor evolution of the neutrinos is independent of their trajectories, which allows us to replace the angle-dependent interaction coupling with one that only depends on the radius, $r$. The neutrino bulb model then gives an approximation for this $r$ dependence, as we expand upon shortly.

The Hamiltonian for neutrino-neutrino interactions at $n$ evenly-spaced discrete momenta in the single-angle approximation is~\cite{Pehlivan:2014zua,siwach:2023prd}
\begin{equation} \label{eq:Ham-3fl}
    H = \sum_{q=1}^n q \vec{B} \cdot \vec{Q}_q + \mu (t) \sum_{q < q'} \vec{Q}_q \cdot \vec{Q}_{q'} \,,
\end{equation}
where $(Q_q)_i = \frac{1}{2} \sum_{j,k=1}^3 a^\dagger_j (\lambda_{i,q})_{jk} a_k$ are the generators of SU(3) in terms of the Gell-Mann matrices $\lambda_i$ on the $q$th qutrit and $\vec{B} = \{0,0,\omega,0,0,0,0,\frac{2}{\sqrt{3}}\Omega\}$, with
\begin{equation} \label{eq:omegap}
    \omega = -\frac{\delta m^2}{2E} \quad, \quad \Omega = -\frac{\Delta m^2}{2E} \,.
\end{equation}
Here $\delta m^2 = |m_2^2 - m_1^2|$ and $\Delta m^2 \approx |m_3^2 - m_1^2| \approx |m_3^2 - m_2^2|$ are the mass-squared differences of the neutrino mass states and $E$ is the average neutrino energy. The time-dependent coupling strength $\mu(t)$ is determined by the aforementioned neutrino bulb model:
\begin{equation}
    \mu(t) = \mu_0 \left( 1 - \sqrt{1-\left(\frac{R_\nu}{r(t)}\right)^2} \right)^2 \,;
\end{equation}
assuming that neutrinos travel at the speed of light, $r(t) = r_0 + c\,t$. {\color{black} Here $r(t)$ is the distance from the center of the neutrinosphere, and $R_\nu$ is the radius of the neutrinosphere. Properties of this simplified model such as entanglement, spectral splitting, and flavor oscillation have been extensively studied to date~\cite{Martin:2023ljq, cervia:2019, Birol:2018qhx, siwach:2025, siwach:2023prd}.}

Our goal is now mathematically simple and quite familiar: given an initial state $|\psi_0\rangle$, what is the state at a later time $t$, $|\psi(t)\rangle$? How can we best leverage quantum computing's strengths to achieve this?

\section{Hybrid Qutrit Algorithm} \label{sec:qutrit_algo}

As mentioned in the introduction, NISQ devices struggle with entangling gate fidelity, forcing researchers to explore ways to limit excessive use of two-qudit gates in algorithms~\cite{McKay:2017rej,Tiwari:2026qls,Huang2019,Pfeiffer:2026aeq}. This effectively leads to circuit depth minimization; we would rather run many short high-fidelity circuits than run a few long circuits that produce untrustworthy results. This decision can be validated in the CCSN neutrino problem. If we want to perform a simple first-order Trotterization for some Hamiltonian $H$ on a quantum computer, we must find a quantum circuit for the time evolution operator $\exp [-i H \delta t]$. This could be done with an algorithm such as time evolution block decimation (TEBD), where the diagonalization circuit for each block will require two-qubit gates~\cite{TEBD:2003}. The number of blocks (and consequently the number of diagonalization circuits required) increases with the number of neutrinos; even for modestly-sized systems, the required number of entangling gates can be on the order of ten to 100 per time step. With current entangling gate fidelities hovering around 99\%, these circuits are untenable~\cite{Huang2019,Pfeiffer:2026aeq}.

For long time integrations of quantum systems, then, other methods are required in the near term. In this paper, we explore the feasibility of a quantum-assisted simulator (QAS) based on the Dirac-Frenkel variational principle for the time evolution of the supernova neutrino system laid out in Sec.~\ref{sec:supernova}. As we will show, a QAS has the advantage of performing the time evolution classically, leveraging classical computers' efficiency at matrix multiplication. This leaves the task of calculating the time evolution operator matrix elements to the quantum computer.

\subsection{Unitary form of Hamiltonian} \label{sec:hamiltonian}
To find a quantum circuit for our neutrino Hamiltonian $H$ in Eq.~\ref{eq:Ham-3fl}, it is natural to use qutrits, which transform in SU(3) identically to neutrino three-flavor states. Unfortunately, unlike the generators of SU(2) (the Pauli matrices), the Gell-Mann matrices are not unitary and cannot be used as gates on qutrits. We thus need to rewrite the Hamiltonian in terms of unitary matrices. In analogy with SU(2), we can define generalized $X$ and $Z$ gates for qutrits \cite{e15062340,Cui:2014,zxxx}
\begin{equation}
    X = \left( \begin{array}{ccc}
        0 & 0 & 1 \\
        1 & 0 & 0 \\
        0 & 1 & 0
    \end{array} \right) \quad, \quad
    Z = \left( \begin{array}{ccc}
        1 & 0 & 0 \\
        0 & \beta & 0 \\
        0 & 0 & \beta^2
    \end{array} \right)\,,
\end{equation}
where $\beta = \exp{2 \pi i /3}$ is a third root of unity.
The $X$ gate simply cycles the basis states $|0\rangle \rightarrow |1\rangle \rightarrow |2\rangle$, while the $Z$ gate adds phases to the $|1\rangle$ and $|2\rangle$ states. It is straightforward to show that these matrices are unitary, and, using the identity $1 + \beta + \beta^2 = 0$, we see that both are traceless. We can then find a set of operators $\vec{\Sigma}$ composed of products of $X$ and $Z$ that are in one-to-one correspondence with the Gell-Mann matrices:
\begin{equation*}
    \vec{\Sigma} = \{ X,Z,X^2,\beta XZ,Z^2,\beta^2 X Z^2,X^2 Z,X^2 Z^2\}\,.
\end{equation*}
Expressions converting the Gell-Mann matrices to these unitary operators and vice versa can be found in Appendix~\ref{app:mat_translation}.

We can now rewrite the Hamiltonian in terms of matrices in $\vec{\Sigma}$. Writing $H = H_\nu + \mu(t) H_{\nu \nu}$ and defining $\omega_q = q \omega$ and $\Omega_q = q \Omega$, with $\omega$ and $\Omega$ defined in Eq.~\ref{eq:omegap}, we {\color{black}  derive}
{\color{black} \begin{equation} \label{eq:Hnu}
    H_\nu = \sum_{q=1}^n q \left( \varepsilon Z_q + \varepsilon^* Z_q^2 \right) \,,
\end{equation}}
where 
{\color{black} \begin{equation*}
    \varepsilon = \omega \frac{1-\beta^2}{6} - \Omega \frac{\beta}{3} \,.
\end{equation*}}
{\color{black} We calculate the interaction term as}
\begin{equation} \label{eq:Hnunu}
\begin{split}
    H_{\nu\nu} &\equiv \sum_{q < q'} \vec{\lambda}_q \cdot \vec{\lambda}_{q'} = \sum_{q \neq q'} \sum_{i=1}^8 \lambda_{i,q} \otimes \lambda_{i,q'} \\
    &= \sum_{q < q'} \frac{2}{3} [ X_q \otimes X_{q'}^2 + X_q^2 \otimes X_{q'} \\
    &\quad\quad + Z_q \otimes Z_{q'}^2 + Z_q^2 \otimes Z_{q'} \\
    &\quad\quad + \beta\left( X_qZ_q^2 \otimes X_{q'}^2 Z_{q'} + X_q^2 Z_q \otimes X_{q'} Z_{q'}^2 \right) \\
    &\quad\quad + \beta^2 \left( X_q Z_q \otimes X_{q'}^2 Z_{q'}^2 + X_q^2 Z_q^2 \otimes X_{q'} Z_{q'} \right) ] \,,
\end{split}
\end{equation}
where $X_q,Z_q$ act on the $q$th neutrino and the tensor products are assumed to also contain identity operators for all other momentum states $k \neq q, q'$. {\color{black} It is not immediate that $H_{\nu\nu}$ is Hermitian, but using the relations $XZ=\beta^2 ZX$ and $\beta^* = \beta^2$ this can be verified.} For $n$ neutrinos, there will be $8 \cdot \frac{n(n-1)}{2} = 4n(n-1)$ unitary operators in $H_{\nu\nu}$. {\color{black} These operators also have trivial circuits composed only of single-qutrit gates (e.g., $X \otimes X^2 = (X \otimes I) \cdot (I \otimes X^2)$), simplifying the forthcoming calculations.}

\subsection{Dirac-Frenkel time evolution} \label{sec:timeevo}

Time evolution of our neutrino system on a quantum computer would normally follow some form of Trotterization, where the state vector is updated over some time step $\delta t$ by the application of a time evolution operator circuit, $e^{-i H \delta t}$. This often leads to very deep circuits with many entangling gates for even modest numbers of neutrinos and time steps~\cite{Turro:2024, Spagnoli:2025prd}. We employ a different method to avoid this problem, instead using a quantum-assisted simulator based on the Dirac and Frenkel variational principle~\cite{Yuan2019}. Below we provide a short explanation of the procedure; for a lengthier discussion, see Ref.~\cite{Bharti:2021,Siwach:2023wzy}.

Given a state of our system at time $t$, $|\psi(t)\rangle$, we wish to calculate the state at some later $t + \delta t$. We first introduce an ansatz basis of states $\{|\phi_i\rangle\}_i$ and corresponding variational parameters $\{\alpha_i(t) \in \mathbb{C}\}_i$ that accurately represent the state $|\psi(t)\rangle$:
\begin{equation}
    |\psi(t)\rangle = \sum_i \alpha_i(t) |\phi_i\rangle \equiv |\phi(t)\rangle \,.
\end{equation}
The time-evolved state $|\psi(t + \delta t)\rangle$ may not in general be in the vector space of the ansatz basis states, but for small $\delta t$ the approximation is good:
\begin{equation}
    |\psi(t+\delta t)\rangle \approx |\phi(t+\delta t)\rangle \approx |\phi(t)\rangle - i \delta t H |\phi(t)\rangle \,.
\end{equation}
On the other hand, we can expand $|\phi(t + \delta t)\rangle$ in powers of $\delta \alpha_i$ to obtain
\begin{equation}
\begin{split}
    |\phi(t+\delta t)\rangle &\approx |\phi(t)\rangle + \sum_i \frac{\partial |\phi(t)\rangle}{\partial \alpha_i} \delta \alpha_i \\
    &= |\phi(t)\rangle + \sum_i \delta \alpha_i(t) |\phi_i\rangle \,.
\end{split}
\end{equation}
Thus, we need to equate $-i\,\delta t\,H |\phi(t)\rangle$ and $\sum_i \delta \alpha_i(t) |\phi_i\rangle$, which can be achieved by projecting directly onto the ansatz space:
\begin{equation}
    \left(\sum_i \delta \alpha_i^\dagger(t) \langle \phi_i | \right) \left( \frac{d}{dt} + iH \right) \left| \phi(t) \right \rangle = 0 \,.
\end{equation}
This leads to a matrix equation for the time derivatives of the $\alpha_i$ parameters:
\begin{equation} \label{eq:DFevo}
    \mathcal{E} \dot{\vec{\alpha}}(t) = -i\,\mathcal{D} \vec{\alpha}(t) \,, 
\end{equation}
where $\vec{\alpha} = \{ \alpha_i \}_i$ and
\begin{equation*}
    \mathcal{E}_{ij} = \langle \phi_i | \phi_j \rangle \,,\quad \mathcal{D}_{ij} = \langle \phi_i | H | \phi_j \rangle \,,
\end{equation*}
along with the unitarity condition $\vec{\alpha}^\dagger \mathcal{E} \vec{\alpha} = 1$.
Thus, our dynamical variables are now the variational parameters $\alpha_i(t)$, and the resulting differential equation can be easily solved classically. Of course, we are left with the tasks of choosing an ansatz basis and finding the matrix elements of $\mathcal{E}$ and $\mathcal{D}$; we achieve the latter by performing Hadamard tests on a quantum computer.

\subsection{Qutrit Hadamard test} \label{sec:hadamardtest}

We can decompose the matrix elements of $\mathcal{D}$ by noticing that the terms in our Hamiltonian given in Eqs. \ref{eq:Hnu} and \ref{eq:Hnunu} are sums of unitary operators, i.e. $H$ is of the form
\begin{equation}
    H = \sum_k \beta_k U_k + \mu(t) \sum_l \gamma_l V_l \,.
\end{equation}
Thus, we can write $\mathcal{D}$ as time-independent and time-dependent parts, $\mathcal{D} = \mathcal{D}^I + \mu(t) \mathcal{D}^D$ with
\begin{equation}
    \mathcal{D}^I_{ij} = \sum_k \beta_k \langle \phi_i | U_k | \phi_j \rangle \,,\quad \mathcal{D}^D_{ij} = \sum_l \gamma_l \langle \phi_i | V_l | \phi_j \rangle \,.
\end{equation}

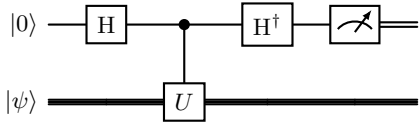
\begin{figure}
    \centering
    \begin{quantikz}[wire types={q,b}]
        \lstick{\ket{0}} & \gate{\text{H}} & \ctrl{1} & \gate{\text{H}^\dagger} & \meter{} & \setwiretype{c} \\
        \lstick{\ket{\psi}}&& \gate{U} &&&
    \end{quantikz}
    \caption{Qutrit Hadamard test to calculate $\text{Re}\langle \psi | U | \psi \rangle$. The expectation value is derived from the measurement probabilities on the ancillary qutrit; see text for the exact equation. Note that the qutrit Hadamard gate is not Hermitian ($\text{H} \neq \text{H}^\dagger$).}
    \label{fig:Hadamard_test_real}
\end{figure}

\begin{figure}
    \centering
    \begin{quantikz}[wire types={q,b}]
        \lstick{\ket{0}} & \gate{\text{H}} & \gate{\text{S}^\dagger} & \ctrl{1} & \gate{\text{H}^\dagger} & \meter{} & \setwiretype{c} \\
        \lstick{\ket{\psi}}&&& \gate{U} &&&
    \end{quantikz}
    \caption{Qutrit Hadamard test to calculate $\text{Im}\langle \psi | U | \psi \rangle$.}
    \label{fig:Hadamard_test_imag}
\end{figure}
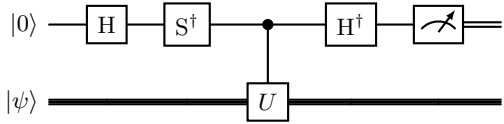

In order to calculate both $\mathcal{E}$ and $\mathcal{D}$, we must calculate expectation values of the form $\langle \phi_i | U | \phi_j \rangle$ for some unitary operator $U$ (in the case of $\mathcal{E}$, the operator is simply the identity matrix). This can be achieved using the well-known Hadamard test, which calculates expectation values of unitary gates $\langle \psi | U | \psi \rangle$, but for qutrits instead of qubits~\cite{quant-algos}. Fig.~\ref{fig:Hadamard_test_real} shows the qutrit circuit to calculate the real part of the expectation value using the qutrit Hadamard gate,
\begin{equation*}
    \text{H} = \frac{1}{\sqrt{3}} \left( \begin{array}{ccc}
        1 & 1 & 1 \\
        1 & \beta & \beta^2 \\
        1 & \beta^2 & \beta
    \end{array} \right)  \,.
\end{equation*}
$\text{Re}\langle \psi | U | \psi \rangle$ is derived from the probability of measuring the three computational basis states of the ancillary qutrit:
\begin{align*}
    P(0) &= \frac{1}{9} \left[ 5 + 2 \langle\psi| (U^{\dagger} + U) |\psi\rangle  \right] \,, \\
    P(1) &= \frac{1}{9} \left[ 2 - \langle\psi| (U^{\dagger} + U) |\psi\rangle \right] \,, \\
    P(2) &= P(1) \,.
\end{align*}
Thus,
\begin{equation}
\begin{split}
    \text{Re}\langle\psi| U |\psi\rangle &= \frac{1}{2} \langle\psi| (U^{\dagger} + U) |\psi\rangle \\
    &= P(0) - \frac{5}{4} \left(P(1) + P(2)\right) \,.
\end{split}
\end{equation}
The imaginary part of $\langle \psi | U | \psi \rangle$ can be calculated with a similar circuit, shown in Fig.~\ref{fig:Hadamard_test_imag}. This circuit uses the Hermitian conjugate of the qutrit S gate,
\begin{equation*}
    \text{S} = \left( \begin{array}{ccc}
        1 & 0 & 0 \\
        0 & \sqrt{\beta} & 0 \\
        0 & 0 & \beta
    \end{array} \right) \,.
\end{equation*}
It can be shown straightforwardly that
\begin{equation}
    \text{Im}\langle \psi | U | \psi \rangle = \frac{3 \sqrt{3}}{4} \left[ P(0) - P(2) \right] \,.
\end{equation}

Now, we are almost ready to use the Hadamard test circuits to calculate $\mathcal{E}$, $\mathcal{D}^I$, and $\mathcal{D}^D$. We notice, however, that the Hadamard tests assume the same state sandwiching the unitary operator, while our matrices' off-diagonal elements contain different $\langle \phi_i |$ and $| \phi_j \rangle$. Thus, we must express the $|\phi_i\rangle$ in terms of a common state, i.e. $| \phi_i \rangle = A_i |\phi_0 \rangle$ for some unitary matrix $A_i$ and state $|\phi_0\rangle$. Then, to calculate $\langle \phi_i | U | \phi_j \rangle = \langle \phi_0| A_i^\dagger U A_j | \phi_0 \rangle$, we replace the unitary operator $U$ in the Hadamard test with $A_i^\dagger U A_j$. Luckily, our procedure for building the ansatz basis states gives us these $A_i$ directly.

\subsection{Ansatz basis}

The central idea behind building the ansatz basis for the quantum-assisted simulator is to find states that are likely to strongly overlap with the ''true" wavefunction $|\psi(t)\rangle$, given an initial state $|\psi(0)\rangle$. For a time-dependent Hamiltonian such as ours in Eq.~\ref{eq:Ham-3fl}, the time evolution operator $e^{-i H \delta t}$ can be written as a sum of commutators of the Hamiltonian with itself (e.g., in a Magnus expansion~\cite{magnus:1954}). Thus, our initial state $|\psi(0)\rangle$ will be acted on by unitary matrices contained in such operators, so it would be advantageous for our ansatz to be comprised of states of the form
\begin{equation*}
\begin{split}
    |\phi_i \rangle &= A_i | \psi(0) \rangle \,, \\
    A_i &\in \left\{ U \in U(3) : U \in \sum_k c_k C_k, \,\,c_k \in \mathbb{C} \right\} \,, \\
    C_k &\in \{ I, H_\nu, [H_\nu, H_{\nu\nu}], [H_\nu,[H_\nu,H_{\nu\nu}]],... \} \,.
\end{split}
\end{equation*}
{\color{black} Here, $U(3)$ denotes the group of $3\times3$ unitary matrices.} That is, we want the $A_i$ to be unitary operators that are contained somewhere in the time evolution operator. We set $A_0 = I$ so that $|\phi_0\rangle = |\psi(0)\rangle$; it will be this state that will be the initial prepared state in the Hadamard tests discussed in the previous section.

\begin{table}[]
    \centering
    \begin{tabular}{c|c|c|c|c}
        $\omega_0\, (\text{eV})$ & $\mu_0\, (\omega_0)$  & $R_\nu\, (\omega_0^{-1})$ & $r_0\, (\omega_0^{-1})$ & $E$\,(MeV) \\
        \hline
        $10^{-10}$ & $3.62\times10^4$ & $32.2$ & $210.64$ & 10 \\
        \hline
        \hline
        $\theta_{12}$\,(deg) & $\theta_{23}$\,(deg) & $\theta_{13}$\,(deg) & $\delta m^2$\,(eV$^2$) & $\Delta m^2$\,(eV$^2$) \\
        \hline
        33.41 & 49.1 & 8.54 & $7.4\times10^{-5}$ & $2.51\times10^{-3}$ \\
    \end{tabular}
    \caption{Values of quantities used in this analysis. We assume natural units $c = \hbar = 1$, and units are shown in parentheses.}
    \label{tab:placeholder}
\end{table}

\begin{figure*}
    \centering
    \includegraphics[width=0.999\linewidth]{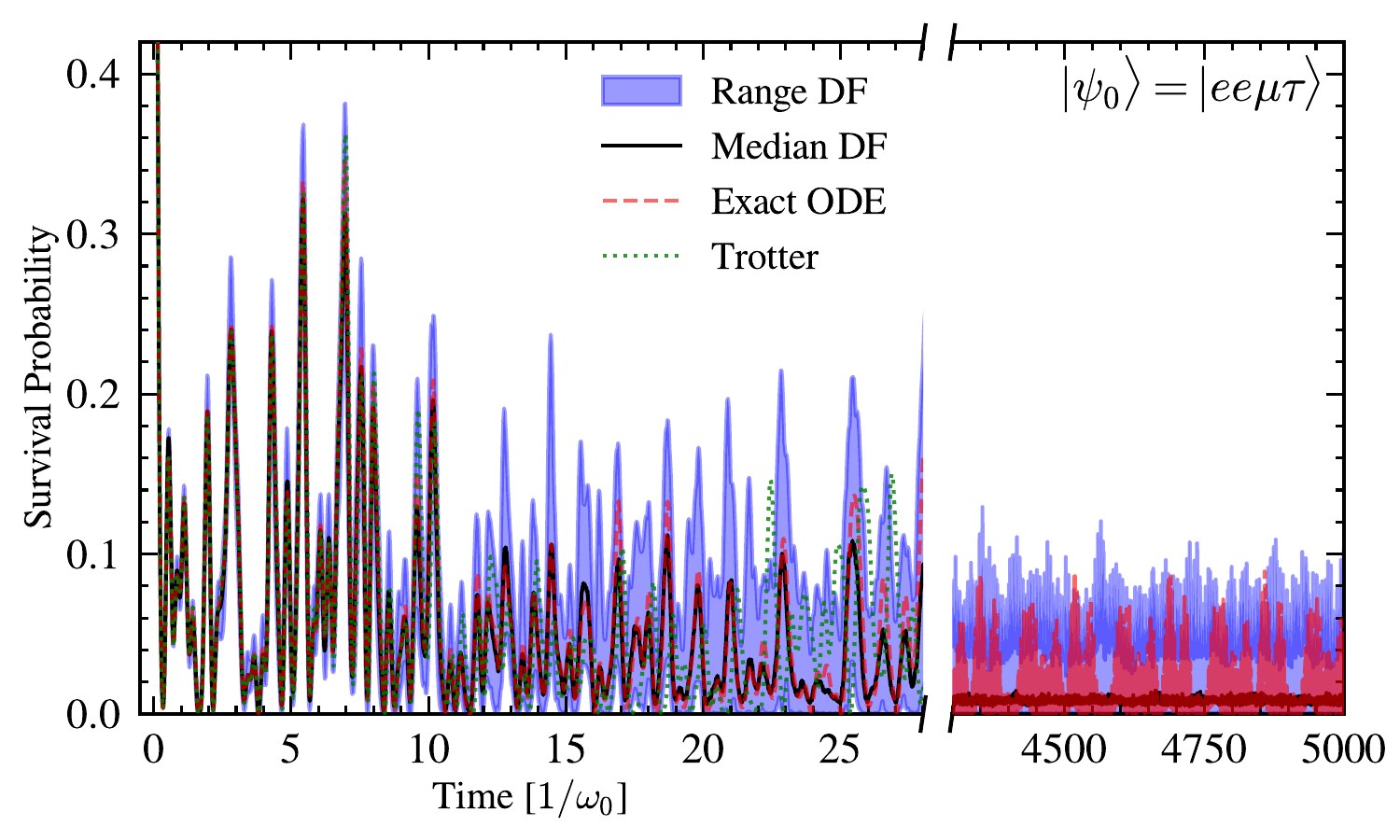}
    \caption{Survival probability of an initial $|e e \mu \tau \rangle$ state, calculated with the Dirac-Frenkel quantum-assisted simulator (median in black and total range in blue), first-order simulated Trotterization (green dotted), and a numerical integration (red dashed). Each Hadamard test in the hybrid simulator used $n_{shots} = 2000000$ shots, and the simulation was repeated $n_{runs} = 65$ times. The Trotter circuit was calculated with $\delta t = 0.005 \, \omega_0$. At late times, the hybrid algorithm loses any resemblance to the exact calculation.}
    \label{fig:eeut_survprob}
\end{figure*}

While this procedure produces a set of states that can be whittled down to a useable basis $\{A_i\}$, we still need to find quantum circuits for each $A_i$ to calculate $\mathcal{E}$ and $\mathcal{D}$. This is achievable given that all the $A_i$ are some combination of commutators of unitary matrices in the Hamiltonian, but procedurally generating them for any number of neutrinos is nontrivial. To simplify this, we choose to define our ansatz basis such that the $A_i$ are easily translated into quantum circuits.

Defining the set $\mathcal{X} = \{ I, X, X^2\}$, we build the $A_i$ by taking every possible tensor product combination of the elements of $\mathcal{X}$ for $n$ neutrinos. {\color{black} Thus, $A_i = X^{i_1} \otimes X^{i_2} \otimes\,...\,\otimes X^{i_n}$, with the exponents $i_j \in \{0,1,2\}$, $j = 1,...,n$. Again, using this procedure $A_0 = I$ ($i_j=0 \,\,\forall \,\,j$). Further, this choice of basis makes the circuits for the $A_i$ trivial (simply applications of $X$ and $X^2$ gates on requisite qutrits).} From our testing, we find this choice of basis does not produce appreciable errors, at least compared to other sources of errors. We also looked at different definitions of $\mathcal{X}$, for example using powers of $XZ$ instead of $X$, and found no significant difference in performance. {\color{black} We note that this ansatz basis is not orthogonal, which is not a particular problem since the Dirac-Frenkel method does not require orthogonality as any overlap in states is taken into account in the $\mathcal{E}_{ij} = \langle \phi_i | \phi_j \rangle$ matrix. However, this does mean that our $3^n$ states do not fully span the entire Hilbert space; by building the basis with $X$ gates applied to our initial state, it is expected that the basis states will have high overlap with the state at time $t$, $|\psi(t)\rangle = e^{-i H t} |\psi(0)\rangle$, given that the time evolution operator contains a sum of commutators of the Hamiltonian with itself (the terms in $C_k$ above).}

\begin{figure*}
    \centering
    \includegraphics[width=0.999\linewidth]{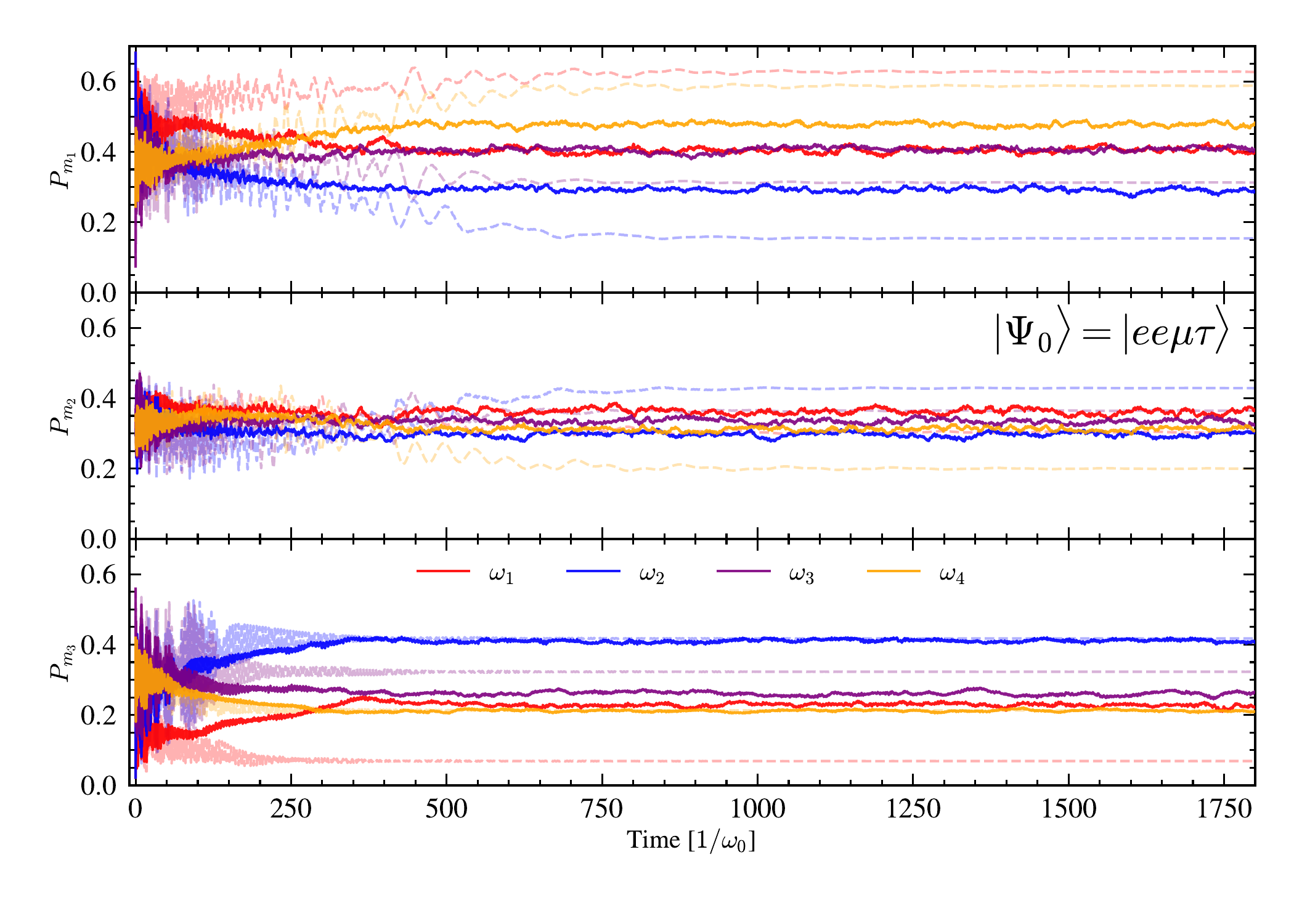}
    \caption{Mass state probability of each particle from an initial $|e e \mu \tau \rangle$ state, calculated with the Dirac-Frenkel quantum-assisted simulator (solid lines) and a numerical integration (dotted lines). Data from the same run as in Fig.~\ref{fig:eeut_survprob}.}
    \label{fig:eeut_massprob}
\end{figure*}

\section{Results} \label{sec:results}

We considered a four neutrino system with three flavors described by Eq.~\ref{eq:Ham-3fl} in the initial state $|ee\mu\tau\rangle$, with the first neutrino having the lowest energy (in this case, the first $\nu_e$). {\color{black} In the following analysis, we use the variable $\omega_0 = 10^{-10}\,$eV as the energy scale of the neutrino vacuum oscillations and redefined dimensionful quantities in terms of this scale.} We calculated the $\mathcal{E}$, $\mathcal{D}^I$, and $\mathcal{D}^D$ matrices by simulating quantum Hadamard test circuits for each matrix element as described in Sec.~\ref{sec:hadamardtest}. We simulated each Hadamard circuit 2,000,000 times using the \texttt{QuForge}\cite{QuForge} qudit simulation package to obtain estimates of the measurement probabilities on the ancillary qutrit, and each matrix element was independently calculated 65 times. Then, for each run, we simulated the time evolution of the $\vec{\alpha}(t)$ (Eq.~\ref{eq:DFevo}) classically with time step $\delta t = 0.005\,\omega_0^{-1}$ out to final time $t_f=5000\,\omega_0^{-1}$ using the \texttt{scipy.integrate.solve\_ivp} method with the DOP853 eighth-order Runge-Kutta method~\cite{2020SciPy-NMeth}. {\color{black} This time step $\delta t$ was chosen to be small enough to fully resolve the flavor oscillations, and we chose $t_f$ such that the mass state probability and entropy reach constant, asymptotic values for each neutrino.} We also solved the Schr{\"o}dinger equation explicity, again using \texttt{scipy.integrate.solve\_ivp} to find $|\psi(t)\rangle$ in the interval $[0,t_f]$. {\color{black} In the following text and figures, the Dirac-Frenkel method is referred to as ``DF" and/or drawn in solid lines, while the numerical Schr{\"o}dinger equation solution is called ``ODE" or ``Exact ODE" and/or drawn in dashed lines.}

Figure~\ref{fig:eeut_survprob} shows the survival probability of our initial $|ee\mu\tau\rangle$ state, {\color{black} defined as $P(\text{surv.)} = |\langle ee\mu\tau| \psi(t)\rangle | ^2$,} at early times (left) and late times (right) calculated with the DF hybrid algorithm. We see that the median of the hybrid algorithm runs (shown in black) matches the numerical solution (red dotted line) fairly well, although as the integration continues, the amplitude of the oscillations becomes damped. This damping is caused by slight variations in the time evolution matrices $\mathcal{E}$ and $\mathcal{D}$ of the 65 runs that are being averaged over, and taken to late times causes a complete loss of any signal. However, each individual run still has nontrivial oscillations in the survival probability, as highlighted by the wide range of hybrid algorithm data (blue band in the figure), even at late times.

{\color{black} Figure~\ref{fig:eeut_massprob} shows the probability of finding each neutrino $q$ from the initial $|ee\mu\tau\rangle$ state in the $i^\text{th}$ mass state $|\nu_i\rangle$ over time, defined as $P_{m_i,q}(t) = \text{tr}[\rho_q(t)\Pi_{i,q}]$ where $\Pi_{i,q} = \left(\bigotimes_{k \neq q} \mathbb{1}_k \right) \otimes |\nu_{i}\rangle\langle\nu_{i}|_q$ and $\rho_q$ is the reduced density matrix for the $q^\text{th}$ neutrino.} We see a similar phenomenon occurring here as in Fig.~\ref{fig:eeut_survprob}, where the hybrid algorithm slowly deviates from the exact solution (dotted lines) due to small deviations in the time evolution matrices. By the time we reach the asymptotic region, the data from the various hybrid algorithm runs have been smoothed out and no longer resembles the expected behavior.

\begin{figure*}
    \centering
    \includegraphics[width=0.999\linewidth]{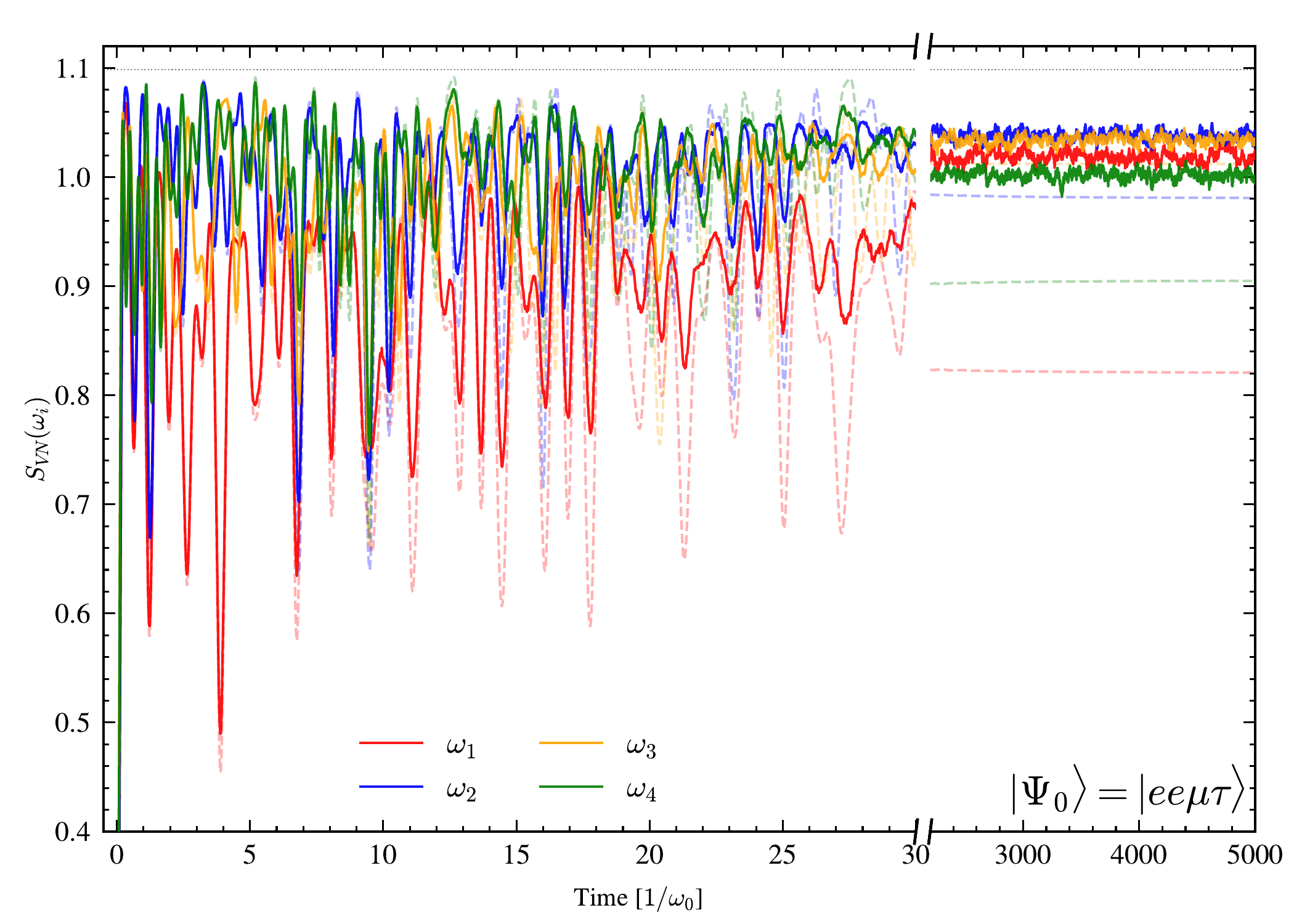}
    \caption{Von Neumann entropy of each particle from an initial $|e e \mu \tau \rangle$ state, calculated with the Dirac-Frenkel quantum-assisted simulator (solid lines) and a classical numerical integration (dotted lines). Data from the same run as in Fig.~\ref{fig:eeut_survprob}.}
    \label{fig:eeut_entropy}
\end{figure*}

We again find a similar story in Figure~\ref{fig:eeut_entropy}, which shows the von Neumann entropy of the four neutrinos as a function of time. {\color{black} The von Neumann entropy for the $i^{\text{th}}$ neutrino is defined as
\begin{equation*}
    S_{VN,i}(t) = -\text{tr}(\rho_i \ln \rho_i) \,,
\end{equation*}
where $\rho_i$ is the reduced density matrix for the $i^\text{th}$ neutrino.} The oscillations in entropy become damped over time such that at late times the asymptotic values deviate from the exact solution.


\subsection{Error Analysis}

Despite the deviations from the exact numerical integration results that we saw in the previous section, the Dirac-Frenkel quantum variational method is still useful to avoid common pitfalls in current noisy quantum time evolution simulations. Circuit depth that scales with the number of time steps and Trotterization errors from building circuits for the time evolution operator $e^{-iH\delta t}$ are two barriers that limit quantum computers' effectiveness in the near-term, and the hybrid algorithm avoids both. However, the hybrid algorithm, like any numerical computation, has its own limitations. The dimension of the $\mathcal{E}$ and $\mathcal{D}$ matrices still scales as $3^n$, and the number of unitary operators in the Hamiltonian increases with the system size as well. Thus, the number of Hadamard tests that need to be performed increases faster than $3^n$, requiring much more quantum computing time for larger systems.

Another issue that arose during our study was the error of $\mathcal{E}$ and $\mathcal{D}$ matrix elements due to the probabilistic nature of quantum measurement. The $\alpha$-norm is defined as $\vec{\alpha}^\dagger \mathcal{E} \vec{\alpha}$ is equal to one assuming $\mathcal{E}$ and $\mathcal{D}$ are Hermitian. {\color{black} To quantify the effect of matrix errors on the simulation results, we performed the following analysis: (1) the $\vec{\alpha}$-vector was calculated exactly using a classical numerical method, denoted $\vec{\alpha}_{cl}$; (2) ``noisy" $\mathcal{E}$ and $\mathcal{D}$ matrices were produced by injecting random complex-valued Gaussian noise into each matrix element of the classically-calculated matrices for multiple different magnitudes of noise; (3) $\vec{\alpha}(t)$ was calculated using the noisy evolution matrices by the same numerical integration used from the Dirac-Frenkel method in the main analysis of this paper; (4) the overlap $\vec{\alpha}_{cl}^\dagger\, \mathcal{E}_{cl} \,\vec{\alpha}(t)$ was calculated, where $\mathcal{E}_{cl}$ is the $\mathcal{E}$ with no injected noise. Crucially, we maintain the Hermiticity of the noisy $\mathcal{E}$ and $\mathcal{D}$ matrices by producing random noise only for the upper triangular portion of the matrix and adding the complex conjugate of that noise to the lower triangular part. This matches how we calculate these matrices in the Dirac-Frenkel method, where we assume the matrices are Hermitian and thus only perform Hadamard tests for the upper triangular part of the matrices. (As an example, for two ansatz basis states $|\phi_i\rangle$ and $|\phi_j\rangle$, $(\mathcal{D}_{ij})^\dagger = (\langle \phi_i| H | \phi_j\rangle)^\dagger = \langle \phi_j| H | \phi_i\rangle = \mathcal{D}_{ji}$).}

\begin{figure}
    \centering
    \includegraphics[width=0.99\linewidth]{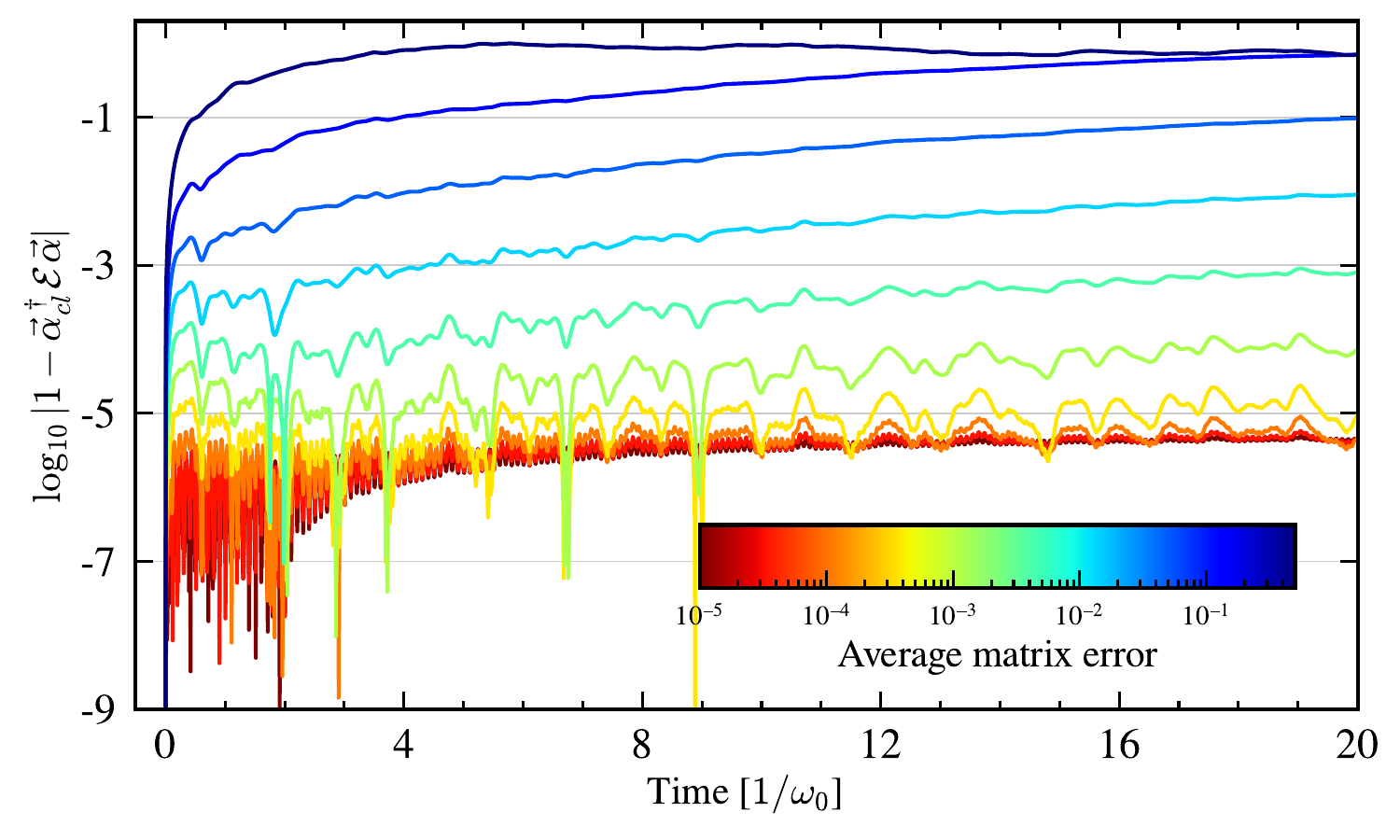}
    \caption{{\color{black} Error in the norm $\vec{\alpha}_{cl}^\dagger \mathcal{E} \vec{\alpha}$ as a function of integration time for different evolution matrix error scales, where $\vec{\alpha}_{cl}$ is the $\alpha$ vectors computed with classically-calculated evolution matrices (as opposed to Hadamard test circuits as done in the main analysis). We take the initial state of the system to be $|ee\mu\tau\rangle$ as in the main analysis. As the error in the matrix elements and integration time increase, so too does the deviation from the classical result.}}
    \label{fig:DF_error}
\end{figure}

{\color{black} The deviation of the noisy values $\vec{\alpha}(t)$ from the noiseless $\vec{\alpha}_{cl}$ can be quantified with the measure $1 - \vec{\alpha}_{cl}^\dagger\, \mathcal{E}_{cl} \,\vec{\alpha}(t)$. In Figure~\ref{fig:DF_error}, we plot the absolute value of this measure as a function of integration time for different magnitudes of injected matrix error. Not surprisingly, the larger the error in the evolution matrices, the larger the deviation from the numerical results. Also, this deviation increases with time, setting an effective maximum time limit for the simulation.}


We see that below a matrix error of $\sim10^{-2}$, the error in the $\alpha$-norm remains below $1\%$ out to time $t \approx 20\,\omega_0^{-1}$, while errors $10^{-2}$ and larger quickly deviate from the classical values (over $10\%$ by $t=20\,\omega_0^{-1}$). Thus, in order to achieve satisfactory results from the Dirac-Frenkel hybrid algorithm, errors in the time evolution matrices must be kept very small. The simplest method to decrease these matrix errors is to simply run more shots in the Hadamard tests that calculate the time evolution matrix elements. In our calculations of the four neutrino system above, we achieved matrix errors averaging about $10^{-2}$ with $2 \times 10^6$ shots; this produced results reasonably close to the classical calculation out to times on the order of $10-30\,\omega_0^{-1}$. {\color{black} The addition of gate noise would of course build upon the shot noise from the measurement of the Hadamard circuits. Our analysis here is agnostic to the source of the noise, so we can determine an estimate for the accuracy of a particular simulation given the number of shots per Hadamard test and gate fidelity. For example, a two-qutrit gate fidelity of $99.9\%$ in the Hadamard tests would introduce to the time evolution matrices an error of $\sim1\%$, which according to Fig.~\ref{fig:DF_error} would still keep the simulation accurate to $\sim1\%$ level out to $t=20\,\omega_0^{-1}$ (assuming no shot noise). See Section~\ref{sec:comp_to_trotter} for further discussion of the impact of two-qutrit (entangling) gate fidelity on the performance of the Hadamard tests in the Dirac-Frenkel method.}

\section{Discussion} \label{sec:discussion}

The Dirac-Frenkel quantum-assisted simulator, in the current context of quantum computing, provides a workaround for current NISQ quantum devices: instead of attempting to evaluate a long, deep time evolution circuit, we can perform many short Hadamard test circuits to calculate the time evolution operator which is used for classical time integration. 

Nevertheless, there are many algorithmic and computational hurdles that still exist for the Dirac-Frenkel hybrid simulator, some of which we have highlighted here in this study. Most glaringly, the results from this method depend strongly on how accurately the time evolution matrix elements are calculated by the quantum Hadamard tests. Increased accuracy will allow for longer time integrations, but determining this required accuracy for a given time interval is a nontrivial task. Because of the probabilistic nature of measuring Hadamard tests, better accuracy can be achieved simply by running more shots per circuit, as in the limit $n_{shots} \rightarrow \infty$ the matrix elements should approach the classically calculated values. Thus, unlike errors in other quantum time evolution algorithms such as Trotterization, there is a very clear path to mitigating these matrix element errors--throw more computing time at the problem.

The Dirac-Frenkel hybrid algorithm also does not solve the issue of classically simulating a Hilbert space that scales exponentially with the number of neutrinos. As mentioned in the previous section on error analysis, the number of degrees of freedom of the system scales as $3^n$ which still needs to be time-evolved on a classical computer. The Dirac-Frenkel equations comprise a system of coupled first-order ordinary differential equations, which, while certainly easy to solve in the grand scheme of differential equations, still require considerable compute time and memory as the number of equations increases. For a 20 neutrino system, the matrix dimension would already be over three billion. One way to mitigate this exponential scaling is to cull the size of the ansatz basis to keep only states with amplitudes over some threshold within the first few time steps. Indeed, this is perhaps the greatest strength of this type of algorithm: we are free to choose our ansatz basis as we please, as long as that choice preserves the data of the system up to some tolerance. For our small four neutrino state, we chose not to decrease the size of the basis; as the number of neutrinos increases, it seems almost mandatory to do so to make the problem tractable. We leave the study of strategies for ansatz basis culling to a later date.

{\color{black} \subsection{Comparison to Trotterization}} \label{sec:comp_to_trotter}

As mentioned in the error analysis subsection above, the more pressing scaling issue with this hybrid algorithm is the exponential increase in the number of Hadamard tests required in order to calculate the time evolution matrices $\mathcal{E}$ and $\mathcal{D}$. The number of Hadamard tests needed is equal to $n_{tests} = n_{U} \cdot n_{basis} \cdot n_{shots} \cdot n_{runs}$, where $n_U$ is the number of unitary operators in the Hamiltonian, $n_{basis}$ is the size of the ansatz basis, $n_{shots}$ is the number of shots per circuit, and $n_{runs}$ is the number of independent runs of the system to average over. In our study, with four neutrinos, $n_U = 56$, $n_{shots} = 2\times10^6$, and $n_{runs} = 65$, we had to take $\sim5.8\times10^{11}$ measurements of Hadamard circuits. Of course, simulating these tests on a classical computer allows for some parallelization, but applying this method to quantum hardware does not benefit from this speedup.

Despite these drawbacks, the Dirac-Frenkel hybrid algorithm offers an improvement in time evolution calculations over Trotterization on NISQ devices for several reasons. First, there is no requirement to build a circuit for the time evolution operator $\exp( -iH\delta t)$, which in general is a matrix sum of nested commutators of the Hamiltonian with itself. This almost always requires some form of approximation, for example in time evolution block decimation (TEBD), which introduces errors that scale polynomially with the time step. Second, the hybrid algorithm avoids deep Trotterization circuits that are plagued by unsatisfactory fidelities on entangling gates. While it is true that the Hadamard test also contains entangling gates (since the Hamiltonian unitary operator and ansatz basis operators needs to be controlled by the ancilla qutrit), the number of these gates stays relatively small.

{\color{black} Quantitatively, for our particular system, the maximum number of entangling gates for a Hadamard test of $\mathcal{D}_{ij} \supset \langle \phi_i | U | \phi_j \rangle = \langle \phi_0| A_i^\dagger U A_j | \phi_0 \rangle$ is $12$: four each for the $A_i^\dagger$ and $A_j$ circuits (which are at most four $X$ or $X^2$ gates) and four from one of the unitary operators in the interaction Hamiltonian $H_{\nu\nu}$. In contrast, a first-order Trotterization circuit representing $e^{-i H \delta t}$ also requires $12$ entangling gates. The crucial difference is that any Trotterization error continues to accumulate with each time step, while the Hadamard tests are all independent, short calculations. Also importantly, the size and depth of the Hadamard tests increase linearly with the number of neutrinos (due to the application of the ansatz basis operators $A_i^\dagger$ and $A_j$), since our Hamiltonian only contains unitary operations between at most two neutrinos. In contrast, for a first-order Trotter circuit of the neutrino collective oscillation Hamiltonian, the number of entangling gates required goes as $n(n-1)$ per time step, and Trotter error still scales as $\mathcal{O}(\delta t^2)$. To reiterate, however, the Dirac-Frenkel method achieves this better scaling by requiring many more (shorter) circuits to be run, although these circuits contain very basic gates such as controlled-$X$ and controlled-$Z$ gates.}

\section{Summary \& Conclusions} \label{sec:summary}

We performed a hybrid quantum-classical time evolution simulation based on the Dirac-Frenkel variational principle for a core-collapse supernova neutrino system. We found that for our initial four neutrino coherent flavor state the hybrid algorithm matches the exact numerical result out to times of around $t \approx 20-30\,\omega_0^{-1}$ with time step $\delta t = 0.005 \,\omega_0^{-1}$, where $\omega_0$ is the energy scale of the single neutrino vacuum oscillations. This time resolution over this integration time is currently impossible using a Trotterization circuit on NISQ devices given the state of entangling gate fidelities.

While the Dirac-Frenkel hybrid algorithm has its own drawbacks and exponential scaling issues, it offers a usable and easily-implementable method to study physical systems on current quantum hardware. Further, improvements to the results shown in this paper can be achieved by increasing the computation time theoretically to any precision, as the error in the quantum-calculated time evolution matrices goes to zero as the number of measurement shots goes to infinity. While we hope and expect quantum computers will become more reliable and capable of longer computations in the future, the Dirac-Frenkel hybrid algorithm and quantum-assisted simulators like it are one of the best choices for leveraging quantum hardware today to study physical systems.

\begin{acknowledgments}
This material is based upon work supported in part by the U.S. Department of Energy, Office of Science, Office of Nuclear Physics, under the FRIB Theory Alliance award DE-SC0013617. It was also supported in part by the U.S. National Science Foundation Award PHY-2411495.
\end{acknowledgments}
\appendix

\section{$ZX$ matrices from Gell-Mann matrices} \label{app:mat_translation}

The operator basis for the $ZX$ algebra is
\begin{equation}
    \vec{\Sigma} = \{ \mathbb{1},\,X,\,Z,\,X^2,\,\beta XZ,\, Z^2,\, \beta^2 X Z^2,\,X^2 Z,\, X^2 Z^2\}\,.
\end{equation}
The basis elements can be written as linear combinations of the Gell-Mann matrices (including an ``identity" Gell-Mann matrix $\lambda_0$):
\begin{gather*}
    \mathbb{1} = \lambda_0 \, ,  \\ 
    X = \frac{1}{2} \left[ (\lambda_1 - i \lambda_2) + (\lambda_4 + i \lambda_5) + (\lambda_6 - i \lambda_7) \right] \,, \\
    Z = \frac{1}{2} (1 - \beta) (\lambda_3 + i \lambda_8) \, , \\
    X^2 = \frac{1}{2} \left[ (\lambda_1 + i \lambda_2) + (\lambda_4 - i \lambda_5) + (\lambda_6 + i \lambda_7) \right] \,, \\
    \beta X Z = \frac{1}{2} \left[ \beta (\lambda_1 - i \lambda_2) + (\lambda_4 + i \lambda_5) + \beta^2 (\lambda_6 - i \lambda_7) \right] \, , \\
    Z^2 = \frac{1}{2} (1 - \beta^2) (\lambda_3 - i \lambda_8) \,, \\
    \beta^2 X Z^2 = \frac{1}{2} \left[ \beta^2 (\lambda_1 - i \lambda_2) + (\lambda_4 + i \lambda_5) + \beta (\lambda_6 - i \lambda_7) \right] \, , \\
    X^2 Z = \frac{1}{2} \left[ \beta (\lambda_1 + i \lambda_2) + (\lambda_4 - i \lambda_5) + \beta^2 (\lambda_6 + i \lambda_7) \right] \,, \\
    X^2 Z^2 = \frac{1}{2} \left[ \beta^2 (\lambda_1 + i \lambda_2) + (\lambda_4 - i \lambda_5) + \beta (\lambda_6 + i \lambda_7) \right] \,.
\end{gather*}
Note that these operators can also be written in terms of the $\mathfrak{su}(2)$ subalgebra raising and lowering operators $I^{\pm} = \lambda_1 \pm \lambda_2$, $U^{\pm} = \lambda_4 \pm \lambda_5$, and $V^{\pm} = \lambda_6 \pm \lambda_7$ and two ``number operators'' $N^{\pm} = \lambda_3 \pm i \lambda_8$. It is easy to check that, while these matrices are not Hermitian, they are unitary.

The inverse relations to those above are:
\begin{gather*}
    \lambda_0 = \mathbb{1} \, , \\
    \lambda_1 = \frac{1}{3} \left[ X (\mathbb{1} + Z + Z^2) + X^2 (\mathbb{1} + \beta^2 Z + \beta Z^2) \right] \, , \\
    \lambda_2 = \frac{i}{3} \left[ X (\mathbb{1} + Z + Z^2) - X^2 (\mathbb{1} + \beta^2 Z + \beta Z^2) \right] \, , \\
    \lambda_3 = \frac{1}{3} \left[ (1 - \beta^2) Z + (1 - \beta) Z^2 \right] \, , \\
    \lambda_4 = \frac{1}{3} \left[ X (\mathbb{1} + \beta Z + \beta^2 Z^2) + X^2 (\mathbb{1} + Z + Z^2) \right] \, , \\
    \lambda_5 = \frac{-i}{3} \left[ X (\mathbb{1} + \beta Z + \beta^2 Z^2) - X^2 (\mathbb{1} + Z + Z^2) \right] \, , \\
    \lambda_6 = \frac{1}{3} \left[ X (\mathbb{1} + \beta^2 Z + \beta Z^2) + X^2 (\mathbb{1} + \beta Z + \beta^2 Z^2) \right] \, , \\
    \lambda_7 = \frac{i}{3} \left[ X (\mathbb{1} + \beta^2 Z + \beta Z^2) - X^2 (\mathbb{1} + \beta Z + \beta^2 Z^2) \right] \, , \\
    \lambda_8 = \frac{-1}{\sqrt{3}} \left( \beta Z + \beta^2 Z^2 \right) \, .
\end{gather*}
 These can be easily verified using the definitions of $X$ and $Z$ and the relation $1 + \beta + \beta^2 = 0$. Further, the translation matrix $U$ satisfying $\vec{\Sigma} = U \vec{\lambda}$, where $\vec{\Sigma}$ and $\vec{\lambda}$ are 8-vectors of the ZX basis matrices and Gell-Mann matrices (excluding $\mathbb{1}$ and $\lambda_0$), respectively, is given by
\begin{equation}
    U_{ij} = \frac{1}{2} \text{Tr} (\Sigma_i \lambda_j) \quad .
\end{equation}

\bibliography{hybrid_qutrit}

@article{Spagnoli:2025prd,
  title = {Collective neutrino oscillations in three flavors on qubit and qutrit processors},
  author = {Spagnoli, Luca and Goss, Noah and Roggero, Alessandro and Rrapaj, Ermal and Cervia, Michael J. and Patwardhan, Amol V. and Naik, Ravi K. and Balantekin, A. Baha and Younis, Ed and Santiago, David I. and Siddiqi, Irfan and Aldaihan, Sheakha},
  journal = {Phys. Rev. D},
  volume = {111},
  issue = {10},
  pages = {103054},
  numpages = {21},
  year = {2025},
  month = {May},
  publisher = {American Physical Society},
  doi = {10.1103/gjr1-lf8s},
  url = {https://link.aps.org/doi/10.1103/gjr1-lf8s}
}

@article{Laraib:2025uza,
    author = "Laraib, Zoha and Richers, Sherwood",
    title = "{Many-body simulations of the fast flavor instability}",
    eprint = "2507.02040",
    archivePrefix = "arXiv",
    primaryClass = "astro-ph.HE",
    doi = "10.1103/d6w8-7j9s",
    journal = "Phys. Rev. D",
    volume = "112",
    number = "10",
    pages = "L101304",
    year = "2025"
}

@article{Laraib:2025ziz,
    author = "Laraib, Zoha and Richers, Sherwood",
    title = "{Two-beam multiparticle many-body simulations of inhomogeneous fast flavor instabilities}",
    eprint = "2511.16506",
    archivePrefix = "arXiv",
    primaryClass = "astro-ph.HE",
    doi = "10.1103/jdzt-jhy2",
    journal = "Phys. Rev. D",
    volume = "112",
    number = "12",
    pages = "123045",
    year = "2025"
}

@article{Chernyshev:2024pqy,
    author = "Chernyshev, Ivan and Robin, Caroline E. P. and Savage, Martin J.",
    title = "{Quantum magic and computational complexity in the neutrino sector}",
    eprint = "2411.04203",
    archivePrefix = "arXiv",
    primaryClass = "quant-ph",
    reportNumber = "IQuS@UW-21-091",
    doi = "10.1103/PhysRevResearch.7.023228",
    journal = "Phys. Rev. Res.",
    volume = "7",
    number = "2",
    pages = "023228",
    year = "2025"
}

@article{Bhaskar:2024myw,
    author = "Bhaskar, Ramya and Roggero, Alessandro and Savage, Martin J.",
    title = "{Timescales in many-body fast-neutrino-flavor conversion}",
    doi = "10.1103/PhysRevC.110.045801",
    journal = "Phys. Rev. C",
    volume = "110",
    number = "4",
    pages = "045801",
    year = "2024"
}

@article{Neill:2024klc,
    author = "Neill, Duff and Liu, Hanqing and Martin, Joshua and Roggero, Alessandro",
    title = "{Scattering neutrinos, spin models, and permutations}",
    eprint = "2406.18677",
    archivePrefix = "arXiv",
    primaryClass = "hep-ph",
    reportNumber = "LA-UR-24-25341",
    doi = "10.1103/PhysRevResearch.7.023157",
    journal = "Phys. Rev. Res.",
    volume = "7",
    number = "2",
    pages = "023157",
    year = "2025"
}

@unpublished{Martin:2023ljq,
    author = "Martin, Joshua D. and Roggero, A. and Duan, Huaiyu and Carlson, J.",
    title = "{Many-body neutrino flavor entanglement in a simple dynamic model}",
    eprint = "2301.07049",
    archivePrefix = "arXiv",
    primaryClass = "hep-ph",
    reportNumber = "LA-UR-23-20394",
    month = "1",
    year = "2023"
}

@article{Illa:2022zgu,
    author = "Illa, Marc and Savage, Martin J.",
    title = "{Multi-Neutrino Entanglement and Correlations in Dense Neutrino Systems}",
    eprint = "2210.08656",
    archivePrefix = "arXiv",
    primaryClass = "nucl-th",
    reportNumber = "IQuS@UW-21-034",
    doi = "10.1103/PhysRevLett.130.221003",
    journal = "Phys. Rev. Lett.",
    volume = "130",
    number = "22",
    pages = "221003",
    year = "2023"
}

@unpublished{Kiss:2025jgt,
    author = "Kiss, Oriel and Tavernelli, Ivano and Tacchino, Francesco and Lacroix, Denis and Roggero, Alessandro",
    title = "{Neutrino thermalization via randomization on a quantum processor}",
    eprint = "2510.24841",
    archivePrefix = "arXiv",
    primaryClass = "quant-ph",
    month = "10",
    year = "2025"
}

@article{Turro:2024shh,
    author = "Turro, Francesco and Chernyshev, Ivan A. and Bhaskar, Ramya and Illa, Marc",
    title = "{Qutrit and qubit circuits for three-flavor collective neutrino oscillations}",
    eprint = "2407.13914",
    archivePrefix = "arXiv",
    primaryClass = "quant-ph",
    reportNumber = "IQuS@UW-21-082",
    doi = "10.1103/PhysRevD.111.043038",
    journal = "Phys. Rev. D",
    volume = "111",
    number = "4",
    pages = "043038",
    year = "2025"
}

@unpublished{Carlson:2026mir,
    author = "Carlson, Joseph and Roggero, Alessandro and Neill, Duff",
    title = "{Neutrino Flavor Evolution in High Flux Astrophysical Environments}",
    eprint = "2603.12192",
    archivePrefix = "arXiv",
    primaryClass = "hep-ph",
    reportNumber = "LA-UR-25-31666",
    month = "3",
    year = "2026"
}

@article{Lacroix:2024pbb,
    author = "Lacroix, Denis and Bauge, Angel and Yilmaz, Bulent and Mangin-Brinet, Mariane and Roggero, Alessandro and Balantekin, A. Baha",
    title = "{Phase-space methods for neutrino oscillations: Extension to multibeams}",
    eprint = "2409.20215",
    archivePrefix = "arXiv",
    primaryClass = "hep-ph",
    doi = "10.1103/PhysRevD.110.103027",
    journal = "Phys. Rev. D",
    volume = "110",
    number = "10",
    pages = "103027",
    year = "2024"
}

@article{Cui:2014,
    author = "Cui, Shawn X. and Wang, Zhenghan",
    title = "{Universal quantum computation with metaplectic anyons}",
    doi = "10.1063/1.4914941",
    journal = "J. Math. Phys.",
    volume = "56",
    number = "3",
    pages = "032202",
    year = "2015"
}

@article{Duan:2010bg,
    author = "Duan, Huaiyu and Fuller, George M. and Qian, Yong-Zhong",
    title = "{Collective Neutrino Oscillations}",
    eprint = "1001.2799",
    archivePrefix = "arXiv",
    primaryClass = "hep-ph",
    reportNumber = "LA-UR-09-08309, INT-PUB-10-001",
    doi = "10.1146/annurev.nucl.012809.104524",
    journal = "Ann. Rev. Nucl. Part. Sci.",
    volume = "60",
    pages = "569--594",
    year = "2010"
}

@article{Chakraborty:2016yeg,
    author = "Chakraborty, Sovan and Hansen, Rasmus and Izaguirre, Ignacio and Raffelt, Georg",
    title = "{Collective neutrino flavor conversion: Recent developments}",
    eprint = "1602.02766",
    archivePrefix = "arXiv",
    primaryClass = "hep-ph",
    doi = "10.1016/j.nuclphysb.2016.02.012",
    journal = "Nucl. Phys. B",
    volume = "908",
    pages = "366--381",
    year = "2016"
}

@article{Tamborra:2020cul,
    author = "Tamborra, Irene and Shalgar, Shashank",
    title = "{New Developments in Flavor Evolution of a Dense Neutrino Gas}",
    eprint = "2011.01948",
    archivePrefix = "arXiv",
    primaryClass = "astro-ph.HE",
    doi = "10.1146/annurev-nucl-102920-050505",
    journal = "Ann. Rev. Nucl. Part. Sci.",
    volume = "71",
    pages = "165--188",
    year = "2021"
}

@article{Volpe:2023met,
    author = "Volpe, M. Cristina",
    title = "{Neutrinos from dense environments: Flavor mechanisms, theoretical approaches, observations, and new directions}",
    eprint = "2301.11814",
    archivePrefix = "arXiv",
    primaryClass = "hep-ph",
    doi = "10.1103/RevModPhys.96.025004",
    journal = "Rev. Mod. Phys.",
    volume = "96",
    number = "2",
    pages = "025004",
    year = "2024"
}

@article{Johns:2025mlm,
    author = "Johns, Lucas and Richers, Sherwood and Wu, Meng-Ru",
    title = "{Neutrino Oscillations in Core-Collapse Supernovae and Neutron Star Mergers}",
    eprint = "2503.05959",
    archivePrefix = "arXiv",
    primaryClass = "astro-ph.HE",
    reportNumber = "LA-UR-25-21809",
    doi = "10.1146/annurev-nucl-121423-100853",
    journal = "Ann. Rev. Nucl. Part. Sci.",
    volume = "75",
    number = "1",
    pages = "399--423",
    year = "2025"
}

@article{Balantekin:2018mpq,
    author = "Balantekin, A. B.",
    title = "{Symmetries and Algebraic Methods in Neutrino Physics}",
    eprint = "1809.02539",
    archivePrefix = "arXiv",
    primaryClass = "hep-ph",
    doi = "10.1088/1361-6471/aae3d8",
    journal = "J. Phys. G",
    volume = "45",
    number = "11",
    pages = "113001",
    year = "2018"
}

@Article{e15062340,
AUTHOR = {Howard, Mark and Brennan, Eoin and Vala, Jiri},
TITLE = {Quantum Contextuality with Stabilizer States},
JOURNAL = {Entropy},
VOLUME = {15},
YEAR = {2013},
NUMBER = {6},
PAGES = {2340--2362},
URL = {https://www.mdpi.com/1099-4300/15/6/2340},
ISSN = {1099-4300},
}

@article{zxxx,
   title={Qutrit ZX-calculus is Complete for Stabilizer Quantum Mechanics},
   volume={266},
   ISSN={2075-2180},
   url={http://dx.doi.org/10.4204/EPTCS.266.3},
   DOI={10.4204/eptcs.266.3},
   journal={Electronic Proceedings in Theoretical Computer Science},
   publisher={Open Publishing Association},
   author={Wang, Quanlong},
   year={2018},
   month=Feb, pages={58–70} }

@article{Balantekin:2024pwc,
    author = "Balantekin, A. B. and Suliga, Anna M.",
    title = "{On the properties of qudits}",
    eprint = "2405.13862",
    archivePrefix = "arXiv",
    primaryClass = "quant-ph",
    reportNumber = "N3AS-24-025",
    doi = "10.1140/epja/s10050-024-01347-x",
    journal = "Eur. Phys. J. A",
    volume = "60",
    number = "6",
    pages = "124",
    year = "2024"
}

@article{Mangin-Brinet:2026prd,
  title = {Three-flavor neutrino oscillations using the phase space approach},
  author = {Mangin-Brinet, Mariane and Bauge, Angel and Lacroix, Denis},
  journal = {Phys. Rev. D},
  volume = {113},
  issue = {3},
  pages = {036026},
  numpages = {13},
  year = {2026},
  month = {Feb},
  publisher = {American Physical Society},
  doi = {10.1103/yr2g-1g2g},
  url = {https://link.aps.org/doi/10.1103/yr2g-1g2g}
}

@article{Cerezo:npj2022,
  author    = {Cerezo, M. and Sharma, Kunal and Arrasmith, Andrew and Coles, Patrick J.},
  title     = {Variational quantum state eigensolver},
  journal   = {npj Quantum Information},
  year      = {2022},
  volume    = {8},
  number    = {1},
  pages     = {113},
  doi       = {10.1038/s41534-022-00611-6},
  url       = {https://doi.org/10.1038/s41534-022-00611-6}
}

@article{Chernyshev:prr2025,
  title = {Quantum magic and computational complexity in the neutrino sector},
  author = {Chernyshev, Ivan and Robin, Caroline E. P. and Savage, Martin J.},
  journal = {Phys. Rev. Res.},
  volume = {7},
  issue = {2},
  pages = {023228},
  numpages = {8},
  year = {2025},
  month = {Jun},
  publisher = {American Physical Society},
  doi = {10.1103/PhysRevResearch.7.023228},
  url = {https://link.aps.org/doi/10.1103/PhysRevResearch.7.023228}
}

@article{Yuan2019,
  doi = {10.22331/q-2019-10-07-191},
  url = {https://doi.org/10.22331/q-2019-10-07-191},
  title = {Theory of variational quantum simulation},
  author = {Yuan, Xiao and Endo, Suguru and Zhao, Qi and Li, Ying and Benjamin, Simon C.},
  journal = {{Quantum}},
  issn = {2521-327X},
  publisher = {{Verein zur F{\"{o}}rderung des Open Access Publizierens in den Quantenwissenschaften}},
  volume = {3},
  pages = {191},
  month = oct,
  year = {2019}
}

@ARTICLE{2020SciPy-NMeth,
  author  = {Virtanen, Pauli and Gommers, Ralf and Oliphant, Travis E. and Haberland, Matt and Reddy, Tyler and Cournapeau, David and Burovski, Evgeni and Peterson, Pearu and Weckesser, Warren and Bright, Jonathan and {van der Walt}, St{\'e}fan J. and Brett, Matthew and Wilson, Joshua and Millman, K. Jarrod and Mayorov, Nikolay and Nelson, Andrew R. J. and Jones, Eric and Kern, Robert and Larson, Eric and Carey, C J and Polat, {\.I}lhan and Feng, Yu and Moore, Eric W. and {VanderPlas}, Jake and Laxalde, Denis and Perktold, Josef and Cimrman, Robert and Henriksen, Ian and Quintero, E. A. and Harris, Charles R. and Archibald, Anne M. and Ribeiro, Ant{\^o}nio H. and Pedregosa, Fabian and {van Mulbregt}, Paul and {SciPy 1.0 Contributors}},
  title   = {{{SciPy} 1.0: Fundamental Algorithms for Scientific Computing in Python}},
  journal = {Nature Methods},
  year    = {2020},
  volume  = {17},
  pages   = {261--272},
  adsurl  = {https://rdcu.be/b08Wh},
  doi     = {10.1038/s41592-019-0686-2}
}

@article{QuForge,
title = {QuForge: A library for qudits simulation},
journal = {Computer Physics Communications},
volume = {314},
pages = {109687},
year = {2025},
issn = {0010-4655},
doi = {https://doi.org/10.1016/j.cpc.2025.109687},
url = {https://www.sciencedirect.com/science/article/pii/S0010465525001894},
author = {Tiago {de Souza Farias} and Lucas Friedrich and Jonas Maziero}
}

@article{Turro:2024,
  title = {Qutrit and qubit circuits for three-flavor collective neutrino oscillations},
  author = {Turro, Francesco and Chernyshev, Ivan A. and Bhaskar, Ramya and Illa, Marc},
  journal = {Phys. Rev. D},
  volume = {111},
  issue = {4},
  pages = {043038},
  numpages = {19},
  year = {2025},
  month = {Feb},
  publisher = {American Physical Society},
  doi = {10.1103/PhysRevD.111.043038},
  url = {https://link.aps.org/doi/10.1103/PhysRevD.111.043038}
}

@article{siwach:2023prd,
  title = {Entanglement in three-flavor collective neutrino oscillations},
  author = {Siwach, Pooja and Suliga, Anna M. and Balantekin, A. Baha},
  journal = {Phys. Rev. D},
  volume = {107},
  issue = {2},
  pages = {023019},
  numpages = {12},
  year = {2023},
  month = {Jan},
  publisher = {American Physical Society},
  doi = {10.1103/PhysRevD.107.023019},
  url = {https://link.aps.org/doi/10.1103/PhysRevD.107.023019}
}

@article{siwach:2025,
  title = {Exploring entanglement and spectral split correlations in three-flavor collective neutrino oscillations},
  author = {Siwach, Pooja and Balantekin, A. Baha and Patwardhan, Amol V. and Suliga, Anna M.},
  journal = {Phys. Rev. D},
  volume = {111},
  issue = {6},
  pages = {063038},
  numpages = {13},
  year = {2025},
  month = {Mar},
  publisher = {American Physical Society},
  doi = {10.1103/PhysRevD.111.063038},
  url = {https://link.aps.org/doi/10.1103/PhysRevD.111.063038}
}

@article{cervia:2019,
  title = {Entanglement and collective flavor oscillations in a dense neutrino gas},
  author = {Cervia, Michael J. and Patwardhan, Amol V. and Balantekin, A. B. and Coppersmith, S. N. and Johnson, Calvin W.},
  journal = {Phys. Rev. D},
  volume = {100},
  issue = {8},
  pages = {083001},
  numpages = {16},
  year = {2019},
  month = {Oct},
  publisher = {American Physical Society},
  doi = {10.1103/PhysRevD.100.083001},
  url = {https://link.aps.org/doi/10.1103/PhysRevD.100.083001}
}

@article{Birol:2018qhx,
    author = "Birol, Savas and Pehlivan, Y. and Balantekin, A. B. and Kajino, T.",
    title = "{Neutrino Spectral Split in the Exact Many Body Formalism}",
    eprint = "1805.11767",
    archivePrefix = "arXiv",
    primaryClass = "astro-ph.HE",
    doi = "10.1103/PhysRevD.98.083002",
    journal = "Phys. Rev. D",
    volume = "98",
    number = "8",
    pages = "083002",
    year = "2018"
}

@article{Amitrano:2022yyn,
  title = {Trapped-ion quantum simulation of collective neutrino oscillations},
  author = {Amitrano, Valentina and Roggero, Alessandro and Luchi, Piero and Turro, Francesco and Vespucci, Luca and Pederiva, Francesco},
  journal = {Phys. Rev. D},
  volume = {107},
  issue = {2},
  pages = {023007},
  numpages = {15},
  year = {2023},
  month = {Jan},
  publisher = {American Physical Society},
  doi = {10.1103/PhysRevD.107.023007},
  url = {https://link.aps.org/doi/10.1103/PhysRevD.107.023007}
}

@article{Yeter-Aydeniz:2021olz,
    author = {Yeter-Aydeniz, K\"ubra and Bangar, Shikha and Siopsis, George and Pooser, Raphael C.},
    title = "{Collective neutrino oscillations on a quantum computer}",
    eprint = "2104.03273",
    archivePrefix = "arXiv",
    primaryClass = "quant-ph",
    doi = "10.1007/s11128-021-03348-x",
    journal = "Quant. Inf. Proc.",
    volume = "21",
    number = "3",
    pages = "84",
    year = "2022"
}

@article{Hall:2021rbv,
    author = "Hall, Benjamin and Roggero, Alessandro and Baroni, Alessandro and Carlson, Joseph",
    title = "{Simulation of collective neutrino oscillations on a quantum computer}",
    eprint = "2102.12556",
    archivePrefix = "arXiv",
    primaryClass = "quant-ph",
    doi = "10.1103/PhysRevD.104.063009",
    journal = "Phys. Rev. D",
    volume = "104",
    number = "6",
    pages = "063009",
    year = "2021"
}

@article{Cervia:2022pro,
    author = "Cervia, Michael J. and Siwach, Pooja and Patwardhan, Amol V. and Balantekin, A. B. and Coppersmith, S. N. and Johnson, Calvin W.",
    title = "{Collective neutrino oscillations with tensor networks using a time-dependent variational principle}",
    eprint = "2202.01865",
    archivePrefix = "arXiv",
    primaryClass = "hep-ph",
    doi = "10.1103/PhysRevD.105.123025",
    journal = "Phys. Rev. D",
    volume = "105",
    number = "12",
    pages = "123025",
    year = "2022"
}

@article{Pehlivan:2014zua,
    author = "Pehlivan, Y. and Balantekin, A. B. and Kajino, Toshitaka",
    title = "{Neutrino Magnetic Moment, CP Violation and Flavor Oscillations in Matter}",
    eprint = "1406.5489",
    archivePrefix = "arXiv",
    primaryClass = "hep-ph",
    doi = "10.1103/PhysRevD.90.065011",
    journal = "Phys. Rev. D",
    volume = "90",
    number = "6",
    pages = "065011",
    year = "2014"
}

@article{Bharti:2021,
  title = {Quantum-assisted simulator},
  author = {Bharti, Kishor and Haug, Tobias},
  journal = {Phys. Rev. A},
  volume = {104},
  issue = {4},
  pages = {042418},
  numpages = {9},
  year = {2021},
  month = {Oct},
  publisher = {American Physical Society},
  doi = {10.1103/PhysRevA.104.042418},
  url = {https://link.aps.org/doi/10.1103/PhysRevA.104.042418}
}

@article{Friedland:2006ke,
    author = "Friedland, Alexander and McKellar, Bruce H. J. and Okuniewicz, Ivona",
    title = "{Construction and analysis of a simplified many-body neutrino model}",
    eprint = "hep-ph/0602016",
    archivePrefix = "arXiv",
    reportNumber = "LA-UR-05-8831",
    doi = "10.1103/PhysRevD.73.093002",
    journal = "Phys. Rev. D",
    volume = "73",
    pages = "093002",
    year = "2006"
}

@article{Bell:2003mg,
    author = "Bell, Nicole F. and Rawlinson, Andrew A. and Sawyer, R. F.",
    title = "{Speedup through entanglement: Many body effects in neutrino processes}",
    eprint = "hep-ph/0304082",
    archivePrefix = "arXiv",
    reportNumber = "FERMILAB-PUB-03-062-A",
    doi = "10.1016/j.physletb.2003.08.035",
    journal = "Phys. Lett. B",
    volume = "573",
    pages = "86--93",
    year = "2003"
}

@article{Lacroix:2022krq,
  title = {Role of non-Gaussian quantum fluctuations in neutrino entanglement},
  author = {Lacroix, Denis and Balantekin, A. B. and Cervia, Michael J. and Patwardhan, Amol V. and Siwach, Pooja},
  journal = {Phys. Rev. D},
  volume = {106},
  issue = {12},
  pages = {123006},
  numpages = {16},
  year = {2022},
  month = {Dec},
  publisher = {American Physical Society},
  doi = {10.1103/PhysRevD.106.123006},
  url = {https://link.aps.org/doi/10.1103/PhysRevD.106.123006}
}

@article{Duan:2006an,
    author = "Duan, Huaiyu and Fuller, George M. and Carlson, J and Qian, Yong-Zhong",
    title = "{Simulation of Coherent Non-Linear Neutrino Flavor Transformation in the Supernova Environment. 1. Correlated Neutrino Trajectories}",
    eprint = "astro-ph/0606616",
    archivePrefix = "arXiv",
    reportNumber = "LA-UR-06-4274",
    doi = "10.1103/PhysRevD.74.105014",
    journal = "Phys. Rev. D",
    volume = "74",
    pages = "105014",
    year = "2006"
}

@article{Qian:1994wh,
    author = "Qian, Yong Zhong and Fuller, George M.",
    title = "{Neutrino-neutrino scattering and matter enhanced neutrino flavor transformation in Supernovae}",
    eprint = "astro-ph/9406073",
    archivePrefix = "arXiv",
    reportNumber = "DOE-ER-40561-150, INT-94-00-63",
    doi = "10.1103/PhysRevD.51.1479",
    journal = "Phys. Rev. D",
    volume = "51",
    pages = "1479--1494",
    year = "1995"
}

@article{Mirizzi:2015eza,
    author = "Mirizzi, Alessandro and Tamborra, Irene and Janka, Hans-Thomas and Saviano, Ninetta and Scholberg, Kate and Bollig, Robert and Hudepohl, Lorenz and Chakraborty, Sovan",
    title = "{Supernova Neutrinos: Production, Oscillations and Detection}",
    eprint = "1508.00785",
    archivePrefix = "arXiv",
    primaryClass = "astro-ph.HE",
    doi = "10.1393/ncr/i2016-10120-8",
    journal = "Riv. Nuovo Cim.",
    volume = "39",
    number = "1-2",
    pages = "1--112",
    year = "2016"
}

@article{Balantekin:2023ayx,
    author = "Balantekin, A. Baha and Cervia, Michael J. and Patwardhan, Amol V. and Surman, Rebecca and Wang, Xilu",
    title = "{Collective Neutrino Oscillations and Heavy-element Nucleosynthesis in Supernovae: Exploring Potential Effects of Many-body Neutrino Correlations}",
    eprint = "2311.02562",
    archivePrefix = "arXiv",
    primaryClass = "astro-ph.HE",
    reportNumber = "N3AS-24-004",
    doi = "10.3847/1538-4357/ad393d",
    journal = "Astrophys. J.",
    volume = "967",
    number = "2",
    pages = "146",
    year = "2024"
}

@article{Siwach:2023wzy,
    author = "Siwach, Pooja and Harrison, Kaytlin and Balantekin, A. Baha",
    title = "{Collective neutrino oscillations on a quantum computer with hybrid quantum-classical algorithm}",
    eprint = "2308.09123",
    archivePrefix = "arXiv",
    primaryClass = "quant-ph",
    reportNumber = "LLNL-JRNL-853275-DRAFT",
    doi = "10.1103/PhysRevD.108.083039",
    journal = "Phys. Rev. D",
    volume = "108",
    number = "8",
    pages = "083039",
    year = "2023"
}

@article{Balantekin:2023qvm,
    author = "Balantekin, A. B. and Cervia, Michael J. and Patwardhan, Amol V. and Rrapaj, Ermal and Siwach, Pooja",
    title = "{Quantum information and quantum simulation of neutrino physics}",
    eprint = "2305.01150",
    archivePrefix = "arXiv",
    primaryClass = "nucl-th",
    doi = "10.1140/epja/s10050-023-01092-7",
    journal = "Eur. Phys. J. A",
    volume = "59",
    number = "8",
    pages = "186",
    year = "2023"
}

@inbook{Patwardhan:2022mxg,
    author = "Patwardhan, Amol V. and Cervia, Michael J. and Rrapaj, Ermal and Siwach, Pooja and Balantekin, A. B.",
    editor = "Tanihata, Isao and Toki, Hiroshi and Kajino, Toshitaka",
    title = "{Many-Body Collective Neutrino Oscillations: Recent Developments}",
    booktitle = "{Handbook of Nuclear Physics}",
    eprint = "2301.00342",
    archivePrefix = "arXiv",
    primaryClass = "hep-ph",
    doi = "10.1007/978-981-15-8818-1_126-1",
    pages = "1--16",
    year = "2023"
}

@article{Pitik:2022,
  title = {Exploiting stellar explosion induced by the QCD phase transition in large-scale neutrino detectors},
  author = {Pitik, Tetyana and Heimsoth, Daniel J. and Suliga, Anna M. and Balantekin, A. Baha},
  journal = {Phys. Rev. D},
  volume = {106},
  issue = {10},
  pages = {103007},
  numpages = {19},
  year = {2022},
  month = {Nov},
  publisher = {American Physical Society},
  doi = {10.1103/PhysRevD.106.103007},
  url = {https://link.aps.org/doi/10.1103/PhysRevD.106.103007}
}

@misc{choi:2025,
      title={Predicted Neutrino Signal Features of Core-Collapse Supernovae}, 
      author={Lyla Choi and Adam Burrows and David Vartanyan},
      year={2025},
      eprint={2503.07531},
      archivePrefix={arXiv},
      primaryClass={astro-ph.HE},
      url={https://arxiv.org/abs/2503.07531}, 
}

@unpublished{Fuller:2022nbn,
    author = "Fuller, G. M. and Haxton, W. C.",
    title = "{Neutrinos in Stellar Astrophysics}",
    eprint = "2208.08050",
    archivePrefix = "arXiv",
    primaryClass = "nucl-th",
    month = "8",
    year = "2022"
}

@Inbook{Martinez-Pinedo:2017ksl,
author="Mart{\'i}nez-Pinedo, Gabriel
and Fischer, Tobias
and Langanke, Karlheinz
and Lohs, Andreas
and Sieverding, Andre
and Wu, Meng-Ru",
editor="Alsabti, Athem W.
and Murdin, Paul",
title="Neutrinos and Their Impact on Core-Collapse Supernova Nucleosynthesis",
bookTitle="Handbook of Supernovae",
year="2017",
publisher="Springer International Publishing",
pages="1805--1841",
isbn="978-3-319-21846-5",
doi="10.1007/978-3-319-21846-5_78",
url="https://doi.org/10.1007/978-3-319-21846-5_78"
}

@article{Martin:2021bri,
    author = "Martin, Joshua D. and Roggero, A. and Duan, Huaiyu and Carlson, J. and Cirigliano, V.",
    title = "{Classical and quantum evolution in a simple coherent neutrino problem}",
    eprint = "2112.12686",
    archivePrefix = "arXiv",
    primaryClass = "hep-ph",
    reportNumber = "IQuS@UW-21-016, LA-UR-21-32145, IQuS@UW-21-016,LA-UR-21-32145",
    doi = "10.1103/PhysRevD.105.083020",
    journal = "Phys. Rev. D",
    volume = "105",
    number = "8",
    pages = "083020",
    year = "2022"
}

@article{Martin:2023gbo,
    author = "Martin, Joshua D. and Neill, Duff and Roggero, A. and Duan, Huaiyu and Carlson, J.",
    title = "{Equilibration of quantum many-body fast neutrino flavor oscillations}",
    eprint = "2307.16793",
    archivePrefix = "arXiv",
    primaryClass = "hep-ph",
    reportNumber = "LA-UR-23-28635",
    doi = "10.1103/PhysRevD.108.123010",
    journal = "Phys. Rev. D",
    volume = "108",
    number = "12",
    pages = "123010",
    year = "2023"
}

@misc{quant-algos,
    title={Quantum algorithms for data analysis},
    note={24 June, 2026},
    howpublished={\url{https://quantumalgorithms.org/}},
    author={Luongo, Alessandro}
}

@article{TEBD:2003,
  title = {Efficient Classical Simulation of Slightly Entangled Quantum Computations},
  author = {Vidal, Guifr\'e},
  journal = {Phys. Rev. Lett.},
  volume = {91},
  issue = {14},
  pages = {147902},
  numpages = {4},
  year = {2003},
  month = {Oct},
  publisher = {American Physical Society},
  doi = {10.1103/PhysRevLett.91.147902},
  url = {https://link.aps.org/doi/10.1103/PhysRevLett.91.147902}
}

@article{magnus:1954,
author = {Magnus, Wilhelm},
title = {On the exponential solution of differential equations for a linear operator},
journal = {Communications on Pure and Applied Mathematics},
volume = {7},
number = {4},
pages = {649-673},
doi = {https://doi.org/10.1002/cpa.3160070404},
url = {https://onlinelibrary.wiley.com/doi/abs/10.1002/cpa.3160070404},
eprint = {https://onlinelibrary.wiley.com/doi/pdf/10.1002/cpa.3160070404},
year = {1954}
}

@Article{Huang2019,
author={Huang, W.
and Yang, C. H.
and Chan, K. W.
and Tanttu, T.
and Hensen, B.
and Leon, R. C. C.
and Fogarty, M. A.
and Hwang, J. C. C.
and Hudson, F. E.
and Itoh, K. M.
and Morello, A.
and Laucht, A.
and Dzurak, A. S.},
title={Fidelity benchmarks for two-qubit gates in silicon},
journal={Nature},
year={2019},
month={May},
day={01},
volume={569},
number={7757},
pages={532-536},
issn={1476-4687},
doi={10.1038/s41586-019-1197-0},
url={https://doi.org/10.1038/s41586-019-1197-0}
}

@unpublished{Pfeiffer:2026aeq,
    author = "Pfeiffer, Frederik and others",
    title = "{A high-fidelity two-qubit gate for multimode superconducting P-mon qubits}",
    eprint = "2606.24772",
    archivePrefix = "arXiv",
    primaryClass = "quant-ph",
    month = "6",
    year = "2026"
}

@article{McKay:2017rej,
    author = "McKay, David C. and Wood, Christopher J. and Sheldon, Sarah and Chow, Jerry M. and Gambetta, Jay M.",
    title = "{Efficient Z gates for quantum computing}",
    eprint = "1612.00858",
    archivePrefix = "arXiv",
    primaryClass = "quant-ph",
    doi = "10.1103/PhysRevA.96.022330",
    journal = "Phys. Rev. A",
    volume = "96",
    number = "2",
    pages = "022330",
    year = "2017"
}

@ubpublished{Tiwari:2026qls,
    author = "Tiwari, Tarush and Sahu, Sudhir K. and Ribeill, Guilhem and Senatore, Michael and LaHaye, Matthew D. and Simmonds, Raymond W. and Campbell, Daniel L. and Kamal, Archana and Ranzani, Leonardo",
    title = "{High-fidelity iSWAP gate with Double Transmon Coupler}",
    eprint = "2604.27080",
    archivePrefix = "arXiv",
    primaryClass = "quant-ph",
    month = "4",
    year = "2026"
}

\end{document}